\documentclass[11pt, a4paper]{article}

\def\desX{\ensuremath\mathcal{X}}
 
\usepackage{amsmath,amsbsy,amsfonts,graphicx,url,multirow}
 \usepackage{amsthm} 
\usepackage{subfig}
 \renewcommand{\v}[1]{\boldsymbol{#1}}
 \newcommand{\var}{\mathop{\mathrm{Var}}\nolimits}

 \newcommand{\tr}{\mathop{\mathit{Tr}}\nolimits}
 \def\E{{\mathbb E}\,}

\usepackage{authblk}
\usepackage[a4paper, total={6.7in, 8.2in}]{geometry}
\begin{document}

 \title{A graph-theoretic framework for algorithmic design of experiments}
\author[1]{Ben M.~Parker\thanks{B.M.Parker@soton.ac.uk}}
\author[2]{Steven G Gilmour}
\author[1]{Vasiliki Koutra}
\affil[1]{Southampton Statistical Sciences Research Institute, University of 
Southampton, UK}
\affil[2]{King's College, London}

\maketitle
 \begin{abstract}
 In this paper, we demonstrate that considering experiments in a graph-theoretic manner allows us to exploit automorphisms of the graph to reduce the number of evaluations of candidate designs for those experiments, and thus find optimal designs faster.

We show that the use of automorphisms for reducing the number of evaluations required of an optimality criterion function is effective on designs where experimental units have a network structure. Moreover, we show that we can take block designs with no apparent network structure, such as one-way blocked experiments, row-column experiments, and crossover designs, and add block nodes to induce a network structure. Considering automorphisms can thus reduce the amount of time it takes to find optimal designs for a wide class of experiments.  

 \end{abstract}
\section{Introduction}

In previous work \cite{Parker2016}, a method for designing 
experiments on social networks was introduced. That paper presented a linear network effects model, which provided a framework for finding the optimal design of experiments when experimental units were connected according to some relationship, which was specified by an adjacency matrix, and treatment effects propagated to experimental units connected to the experimental unit to which the treatments were applied. 

Although in that paper the primary application was the design for experiments on social networks, a variety of examples were given showing how these networks could be useful in many applications, for example in agricultural experiments where experimental units were connected by some spatial relationship, and also in crossover trials, where experimental units were connected by temporal networks. 

In that paper, some methods were introduced to exploit symmetries in the network in order to improve the speed of finding an optimal design. In this current paper, we expand this argument, and show that there is a wide class of experiments that can be reformulated into a problem of design on the network, and that by presenting the problem in this networked form, we can suggest how to modify existing algorithms to allow optimal designs on the original problems to be found more quickly by exploiting automorphisms of the network. We recap the previous literature, and the paper we base this work on (\cite{Parker2016}) in what remains of this section. In Section \ref{sec:Algorithm} we give a brief overview of algorithms used currently for design, and propose a new algorithm for social networks based on the properties of the graph. We give examples of use of the new algorithm in Section \ref{sec:Examples}, before showing how this algorithm is useful in a wide variety of design problems, even when there is no obvious network structure. We conclude briefly in Section \ref{sec:Conclusions}.

\subsection{Previous literature}

Due to the increasing prevalence of (particularly) social networks, the general field of network science, and statistical analysis of networks has shown growth over recent years. \cite{Aral2016} provides a review of the importance of statistical inference and design in social networks, describing that although  the literature on analysis of networks has increased rapidly in recent years, there is limited research on experimental design on networks. Much of the research tends to relate to large-scale properties of networks, for example \cite{Basse2017} draw inspiration from model-assisted survey sampling to provide a detailed view of how treatments might be chosen when there are structured relationships represented as networks between experimental units, specifically under a normal-sum model where the response of a node is governed by the sum of normally distributed random variables, determined by the responses of the node's neighbours.

There have been some attempts to make use of graphs as a tool for finding optimal designs, mostly in the context of block designs. \cite{Wit2005}, for the application of designing efficient experiments for dual-channel microarrays, presents a way of representing experiments with blocks of size two, where when a direct comparison is made between two treatments in a block it is represented as a link in a network. Designs (for the $A_s$-optimality criterion as we refer to it in this paper) were found by simulated annealing for up to $n=20$ nodes (treatments in the representation of \cite{Wit2005}). \cite{Bailey2007} extends this work on microarrays, and presents optimal designs. \cite{Bailey2009} generalise this concept to more general block designs by considering the concurrence graph of a block design where nodes representing treatments are joined by a number of edges equal to the number of blocks in which the treatments appear together. By considering properties of the concurrence graph thus formed, they present some interesting combinatoric parallels, and methods for finding optimal designs for these blocked experiments. \cite{Bailey2011} present some parallels between optimal design criteria and well-known properties of graphs using concurrence graphs, in the context of block designs.

The idea of considering isomorphisms of designs - where two or more designs are equivalent- has been used to reduce the search space for finding optimal designs is established: for example \cite{Ma2001} provide an algorithm for fractional factorial designs which makes use of isomorphisms to reduce the complexity of calculations; \cite{Bulutoglu2008} provide a way of classifying orthogonal arrays by isomorphisms. \cite{Colbourn1981} show that the problem of finding whether block designs are isomorphic is equivalent in complexity to the graph automorphism problem. Perhaps surprisingly, this result does not seem to have been used in a constructive method to find designs, as we do in this paper.

Most of the results in this present paper extend \cite{Parker2016}, where we present a linear network effects model which describes the response of each node as depending on the treatment given directly to that node, and also depends on the treatments given to the neighbours of that node. This is summarised in Section \ref{sec:recap} below. Further recent work in this area follows in \cite{Koutra2017}.

\subsection{Recap of design for networks and notation}\label{sec:recap}
\cite{Parker2016} considers networks of experimental units; a network $G=(N,E)$ is an undirected graph, a collection of nodes $N$ and edges $E \subseteq( N \times N)$ where the nodes represent experimental units on each of which we apply some treatment. The edges represent relationships between the experimental units. 

We assume that if a relationship exists between two experimental units, the
response of each experimental unit is dependent on the treatment applied to the
other according to a linear network effects model which we specify below.

We have $|N|=n$ experimental units, and $m$ treatments. The
relationship between experimental units is
specified by the $n\times n$ adjacency matrix $A$ where $A_{ik}=1$ if and only 
if $i$ and $k$ are
related and $A_{ik}=0$ otherwise. By convention, $A_{ii}=0$. 

We assume there is a ``subject effect'' on the response from experimental unit $i$ of
$\tau_{j}$
when
treatment $j$ is given to that subject, and a ``network effect'' of
$\gamma_{l}$
if
treatment $l$ is given to a connected experimental unit $k$ (if $A_{ik}=1$). We
assume
that each experimental unit receives
exactly one treatment.

We measure the response $Y_i$ for each of our experimental units. Let $t(i)$ be the
treatment applied to experimental unit $i$; then our response is modelled as
\begin{equation}
\label{eq:linearNetworkEffects}
Y_i= \mu+\tau_{t(i)} +\sum_{k=1}^n{A_{ik}\gamma_{t(k)}}+\epsilon_i.
\end{equation}
We call this model the ``Linear Network Effects Model'' (LNEM). We assume that errors $\epsilon_i$  are independent and
identically distributed with mean 0 and constant variance $\sigma^2$.
We assume that we wish to estimate the subject and/or network effects, or some
contrast of them.

We let $\v{u}_j$ be the indicator vector
with $i$-th element equal to 1 when treatment $j$ is applied to subject $i$, and 
otherwise 0. In matrix form,
\begin{align}
\label{eq:matForm}
\E(\v{Y})&=\begin{pmatrix}\v{1}& \v{u}_{1} & \v{u}_{2} & \ldots&
\v{u}_{m} & A\v{u}_{1} & A\v{u}_{2}&\ldots &A\v{u}_{m}
\end{pmatrix}\begin{pmatrix}\mu\\\v{\tau}\\\v{
\gamma}\end{pmatrix}=F\v{\beta},\end{align}
where $F$ is our design matrix, and our
vector of parameters is
\begin{equation}
\label{eq:defineBeta}
\v{\beta}=\left(\mu\quad\v{\tau^{T}}\quad\v{\gamma^T}
\right)^T=(\mu\quad\tau_ { 1 } \ldots\tau_{m}
\quad\gamma_1\ldots\gamma_m)^T.\end{equation}


We calculate the Fisher information matrix as $I=F^TF$, and choose an optimality criterion: we seek to minimise the average variance 
of all pairwise differences of treatment
effects,
$$\frac{2}{m(m-1)}\sum_{j=1}^{m-1}{\sum_{l=j+1
}^m{\var({\widehat{\tau_j-\tau_l})}}}.$$ This is defined to be $A_s$-optimality 
for
estimating the
differences in the treatment effects. To ensure estimability, without loss of generality we set $\tau_m=0$.

\section{Algorithm}
\label{sec:Algorithm}
We recap briefly some general properties of design algorithms, which allow us to put our algorithm in context.
Let the design space, the set of all possible assignments of treatments to experimental units for our experiment, be $\mathcal{X}$. To assess how good a design $x\in\mathcal{X}$ is we calculate the value of some optimality criterion $f(x)$, and seek to find the design(s) which maximise $f(x)$; if our optimality criterion is such that we wish to minimise $f(x)$ without loss of generality we maximise $-f(x)$ instead. That is, we wish to find $x^*=\arg\max f(x)$, which we call an optimal design. Typically $f()$ will be some function of the Fisher information matrix, $I(x)$ ; for example A-optimality, a popular criterion, involves taking the average variance of all unknown parameters, which corresponds to $f(x)=\tr(I^{-1}(x))$, the trace of the inverse of the information matrix. D-optimality, which minimises the joint confidence region for the unknown parameters, can be expressed as $f(x)=\det(I^{-1}(x))$. Computationally, $f(x)$ is normally hard to evaluate. For example, in A-optimality both the calculation of this information matrix and the resulting inversion of it will be computationally hard.

Typically when finding optimal designs, the design space $\mathcal{X}$ which we must search over to find a design may be large. For example, if we have $n$ experimental units which each take values in some set $\mathcal{F}$ then the size of the design space will be $|\mathcal{F}|^n$, and we suffer from the ``curse of dimensionality''. Even if $F$ is the binary set $\{0,1\}$ such that each experimental unit may be assigned one of two treatments, then the size of the design space may be $2^n$. If $F$ is a continuous set (e.g. $\mathcal{R}$ or, a set of equivalent size, the interval $[0,1]$, the design space is (infinitely) bigger.

Ideally, we would evaluate the optimality criterion $f(x)$ at all possible designs $x\in\mathcal{X}$ to find the optimal design $x^{*}=\arg\max_{x\in\mathcal{X}}{f(x)}$, but for large design spaces and/or criteria that are difficult to calculate, computational restrictions may mean we can not do this or at least not compute this in a reasonable time. In some cases, analytic results will enable designs to be found exactly without a search algorithm. However, in general we seek some algorithm that enables us to find the optimal design (or a design which is near optimal). In overview, existing algorithms for design seek to do one or both of the following: 

\begin{itemize}
 \item reduce the complexity of the calculation of the optimality criterion $f(.)$
 \item evaluate a subset of $\mathcal{X}$ such that the overall number of calculations of $f(.)$ is smaller.
\end{itemize}

Many algorithms exist within these categories. The first category includes emulation, used extensively within computer designs, where instead of evaluating a complicated function $f()$, we evaluate an emulator $g()$ , a function which is simpler to evaluate but we believe has similar properties such that $f(x)\approx g(x)$ for all $x \in \mathcal{X}$. Updating formula can also be used in some problems, where we know that $f(x+z)=f(x)+h(z)$ for some easy-to-calculate function $h$. 

The second category of algorithms include space-filling designs, where a representative sample of designs that ``fill'' the space $\desX$ are evaluated. There are also many optimisation algorithms, which use the $t$ previously evaluated designs $x_0,x_1,\ldots x_{t-1}$ and their optimality criterion function evaluations $f(x_0),f(x_1),\ldots,f(x_{t-1})$, to choose which design $x_{t}$ should be evaluated next.  Fedorov exchange algorithms, coordinate exchange algorithms, simulated annealing, particle swarm optimisation, many stochastic search algorithms such as Nelder-Mead, are the subject of research. Some of these algorithms are deterministic, in that the choice of $x_t$ is mandated, others are stochastic in that $x_t$ is chosen randomly. To guard against designs which are local maxima, the initial $x_0$ is often chosen randomly, even if the rest of the algorithm is deterministic, and the algorithm repeated with many random $x_0$ chosen as starting designs.

Much of the literature in experimental design revolves around trying to find better algorithms for particular problems. In this work, we do not seek to find new algorithms, but to know how mapping the optimisation problem to a network domain allows us to vastly reduce the number of designs evaluated in order to allow us to find better designs. We can use many existing algorithms with small modifications in this new network framework.

\subsection{Using networks for reducing the state space}
We show in this section how we can use the network representation of the problem as an advantage when we consider algorithms for design.  \cite{Parker2016} introduced some ideas of how we might improve our algorithms, and we recap these here briefly before expanding these ideas and presenting a new algorithm. 

\subsubsection{Symmetry of labels}
For many optimality criteria we are only interested in differences between treatments, the 
treatment effects themselves are irrelevant, and treatments are equivalent up to
relabelling. For example, designs $\{1,2,2,3,1\}$ and $\{1,3,3,2,1\}$ are equivalent. We can
thus reorder any design so that we only evaluate designs where the first
occurrence of label $j$ must come before the first occurrence of label $j+1$.
Without loss of generality, we can assign treatment 1 to experimental unit 1.

\subsubsection{Symmetry of networks}
\begin{figure}[htp]

\centering
\includegraphics[width=0.45\linewidth]{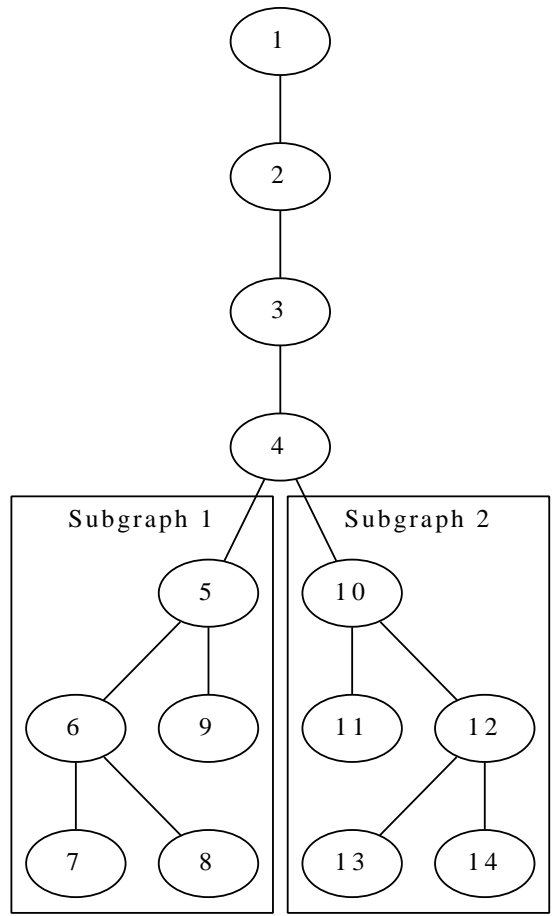}
\caption{\label{fig:automorphism}Example of an automorphism. Subgraphs 1 and 2 can be exchanged, so an automoporphism exists.} 

\end{figure}

In Figure \ref{fig:automorphism}, subgraphs 1 and 2 are exchangeable; i.e. if
we consider subdesign A on subgraph 1, and subdesign B on subgraph 2 (call this
[A1,B2]), we need not also consider [A2,B1] as by symmetry this design has the
same criterion function value. 

We can thus reduce our design space greatly if we can identify subgraphs where
the designs are exchangeable. This is equivalent to finding an
automorphism of our network, a relabelling or permutation of the set $N$
such that the edges $E$ are preserved.  This is known as the Graph Automorphism 
Problem, and finding efficient algorithms to solve this problem has been the topic of much research \cite{Conte2004}. Until recently, the general case of this problem was thought to be not solvable in polynomial time, that is no algorithm existed that could find a solution for a graph with $n$ nodes in $O(n^k)$ time for some constant $k$. (Recently, an algorithm has been found that claims a solution exists in quasi-polynomial  $(\exp((\log n)^{O(1)}))$ time (\cite{Babai2015}, although this result has not yet been peer-reviewed.). 

Whatever the theoretical complexity, in practice fast algorithms exist already that can effectively and quickly find automorphisms in most cases, for example the VF2 algorithm\cite{Cordella2001}, which in the general case finds isomorphisms between two graphs $G_1$ and $G_2$. This algorithm is based on a tree search, where a set of nodes corresponding to a partial match between subgraphs of $G_1$ and $G_2$ is maintained, and then nodes connected to these subgraphs are considered to see if they can extend the matching subgraphs. This algorithm is fast, and implemented by the popular \texttt{igraph} package (\url{http://igraph.org/}) available in many programming languages, including R. Here, we find automorphisms by setting $G_1=G_2$. Many algorithms exist, but we chose VF2 as it was fast enough for our purposes, and was readily incorporated in the software made available with tis paper.

\subsection{New framework for design algorithm}

We present below a general algorithm:
\begin{enumerate}
\item Rewrite original problem in graph form.

\item Find the automorphisms \texttt{isos} of the graph using the VF2 algorithm. This is a set such that for any labelling of the graph corresponding to design $x$, all designs $isos(x)$ will be automorphisms of $x$. 
\item Set $k=1$, $num_{\mbox{eval}}=0$ and pick an initial candidate design $x_1$. 
\begin{enumerate}
\item \label{item:repeat} Check whether $x_k=\min_{\mbox{lex}}{\texttt{isos}(x_k)}$, i.e. if $x_k$ is the first design in the lexicographical order of automorphisms of $x_k$. If so, 
\begin{itemize} 
 \item evaluate the optimality criterion $d_k=f(x_k)$
 \item increment $num_{\mbox{eval}}$ by 1. 
\end{itemize}

\item  Set $k=k+1$. Pick next candidate design $x_k$ according to algorithm of choice \texttt{next} based on $x_1,\ldots,x_{k-1}$ and $d_1,\ldots,d_{k-1}$, so that $x_k=\mbox{next}[(x_1,\ldots,x_{k-1}),(d_1,\ldots,d_{k-1})]$. 
\item If we have reached a stopping criterion, i.e if $\mbox(stop)[(x_1,\ldots,x_k),(d_1,\ldots,d_k),num_{\mbox{eval}}]=1$ for some stopping criterion \texttt{stop}, then stop, otherwise repeat from \ref{item:repeat}.
\end{enumerate}
\end{enumerate}

Typical choices of the \texttt{next} algorithm might be the Fedorov exchange algorithm, an exhaustive search, etc.
Typical choices of the \texttt{stop} criterion might be that $k=k_1$ or $num_{\mbox{eval}}=k_2$ representing some fixed budget on the number of design points examined or function evaluations made, or that $\max (d_{k-t},\ldots,d_k)=d_{k-t}$, i.e. we have not seen an improvement in the last $t$ steps.
\subsubsection{Treatment Structure}
In this paper we restrict ourselves to designs on $m$ unstructured treatments; that is our design space is $\desX=\{(x_1,x_2,\ldots, x_n| x_i\in\{1,\ldots,m\}\forall i\}$ such that $|\desX|=n^m$. We do this for clarity of exposition of our method, but the general ideas extend to factorial and other treatment structures.

\subsubsection{Illustrative example}
Consider the network represented by 3-1-2, where we have three nodes such that nodes 3 and 2 are each connected to node 1 and there are no other connections. It is clear that the network is automorphic to 2-1-3. Let $\mathcal{T}=\{A,B\}$, and let a design be represented by $x_1x_2x_3$ where $x_i\in\mathcal{T}$ is the treatment given to node $i$. In other words $ABB$ means we give treatment A to node 1, B to node 2, and B to node 3. Note that $\mathcal{X}=\{AAA,AAB,ABA,ABB,BAA,BAB,BBA,BBB\}$ where the elements of the design space are clearly in lexicographical order. Let, $X_1=AAA$, and we let \texttt{next} be the next design lexicographically, and let our stopping criterion be \texttt{stop} if $x_k=BBB$.  Then the automorphisms allow us to show that two pairs of designs are equivalent ($AAB=ABA$, $BAB=BBA$); thus in step {\ref{item:repeat}, we do not evaluate $f(x)$ for designs $ABA$ or $BBA$. We have therefore reduced the number of designs we must consider from 8 to 6..

\subsubsection{Rationale\label{sec:rationale}}
For many designs, evaluating the optimality criterion function $f()$ requires (sometimes substantial) computational time, and by reducing the number of evaluations required, we can reduce the computational time required. Alternatively, for the same computational time, we can search more of the design space and hope to find a better design. 

There is some computational overhead in the new algorithm; evaluating the automorphisms initially can be slow, however this needs to be done just once for each network, so for large designs this time is not a large proportion of the total computation time. There is also some overhead in step 3(a), as checking whether a design is lowest in lexicographical ordering of all automorphisms of that design requires some computation. However, we show that this overhead is substantially lower than the cost of evaluating $f()$ in some simple examples, so our algorithm can lead to substantial improvements in finding designs. For particular networks, for example networks with a large number of automorphisms relative to the size of the network, the overhead may be prohibitive. This may be the case for dense or highly regular networks. We investigate the computational complexity of the algorithm in \ref{sec:complexity}.

\section{Examples of use of automorphisms in finding designs}
\label{sec:Examples}
\begin{figure}[htp]
\begin{center}
 \includegraphics[width=0.5\linewidth]{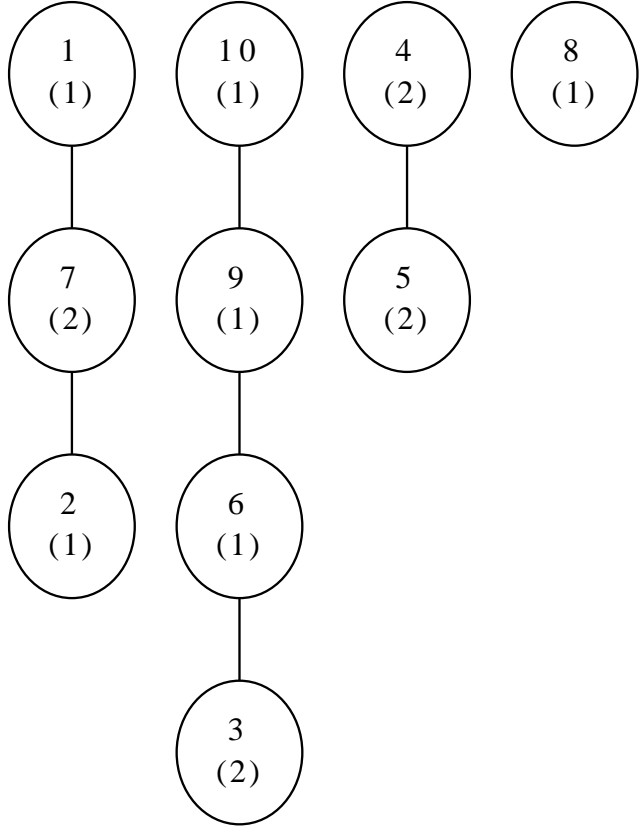}
\end{center}
\caption{\label{fig:JRSSex1}A social network (Example 1 in \cite{Parker2016}. Top nodes are node numbers and bottom nodes are optimal design for $m=2$ treatments for estimating difference in treatment effects allowing for network effects under Linear Network effects model.}
\end{figure}
We return to a small example which was Example 1 in \cite{Parker2016} and represents a social network. This is shown in Figure \ref{fig:JRSSex1}, together with the optimal design for estimating with minimal variance the difference between the $m=2$ treatment effects, $\tau_1-\tau_2$, under the LNEM (\ref{eq:linearNetworkEffects}). The list of edges is shown in the Appendix.

The optimal design was evaluated using exhaustive search. For each candidate design in our design space (there are $\left|\desX\right|  =2^{10}$ designs), the information matrix must be calculated (this a 4 by 4 matrix with columns representing $\mu, \tau_1, \gamma_1, \gamma_2$, recalling we can omit the column corresponding
to $\tau_2=0$). This matrix must be inverted, and the inversion is responsible for much of the computational effort in finding the optimal design. In practice, we can note that as we are only interested in differences between subject effects, we can appeal to symmetry of labels to fix the treatment for experimental unit 1 as the first treatment.

In \cite{Parker2016} we did not use the symmetries of the network, but found the design under exhaustive search; here we use an exhaustive search (so that our \texttt{next} design calculates the lexicographically subsequent design), but will only evaluate designs which are first lexicographically amongst all designs which are automorphisms of the current considered design. We will thus evaluate fewer designs, and \texttt{stop} when one design from each automorphism class has been evaluated.

We naturally find the same optimal design, but evaluate the information matrix 236 times as opposed to 507 times, and use a processing time of 0.02 seconds as opposed to 0.04 seconds.

\begin{table}
\small
\begin{tabular}{p{4cm} l p{2cm}p{2cm}p{2cm}p{2cm}p{2cm}}
  Example &n& Number of automorphisms & Evaluations without automorphisms & Evaluations with automorphisms & Time without automorphisms & Time with automorphisms\\
1. Small social network& 10 & 8 & 507 & 236 & 0.04 & 0.02\\
2. Small social network &  10 & 1 & 511 & 511 & 0.04 & 0.04\\
3. Larger social network & 20 &  8 & 524287 & 221183 & 58.58 & 31.56\\
4. Block design with neighbour effects & 12 & 384 &535008 & 18766 & 108.52 & 33.68\\
5. Non-rectangular field trial& 15 & 2 & 2368741 & 1581572 & 279.6 &197.58\\
6. Crossover trial with dropouts& 15 & 6 & 2262800 & 904555 & 283.86 & 134.26\\ 
\end{tabular}\caption{\label{tab:JRSS2016}Comparison of exhaustive search algorithms with and without automorphisms}
\end{table}

We can repeat this evaluation for all the networks in \cite{Parker2016}, the edge list for each of which is shown in the Appendix. We chose these networks as they represent a variety of different types of network structures, which relate to a variety of different applications. The results are shown in Table \ref{tab:JRSS2016}, and indicate that the number of function evaluations is greatly reduced for all the networks (except network 2, where there are no non-trivial automorphisms) and that the time taken to perform the exhaustive search is greatly reduced, even when taking into account the time needed for calculating automorphisms and lexicographical ordering checking as described above. 

We include the computational time as a guide; this will vary greatly between implementations in different software for these small networks, but gives an idea of how effective computationally the use of automorphisms is for finding designs on network problems, even for small networks.

\subsection{Implementing a simple coordinate exchange algorithm}

We demonstrate how the framework can be combined with a non-trivial design algorithm by showing as an example the cyclic coordinate descent algorithm described fully in\cite{Meyer1995}.

We pick a random starting design. Each node is considered in turn, and for each node we cycle through all possible treatments for that node to see if changing the treatment allocated to that node can improve the design. If we find an improvement, we fix that design, and start again. If after checking all treatments we find that changing the treatment on this node does not improve the design, we move to the next node. When we have cycled through all nodes without improvement, we stop.

In the language of our general algorithm above, we pick
\begin{eqnarray*}x_k&=&\mbox{next}[(x_1,\ldots,x_{k-1}),(d_1,\ldots,d_{k-1})]\\
&=&[x^*_i+u_{k-i,m}]\mod m \end{eqnarray*} 
where $u_{k,m}$ is the unit vector with all elements zero except the $\mbox{floor}(k/m)$-th, which is 1, and $x^{*}_i$ is the maximum design so far discovered such that $x_i^*=\arg\max_{1\le i \le k-1}{f(x_i)}$. for design $i$. The stopping criterion is chosen such that  

$$\mbox{stop}[(x_1,\ldots,x_k),(d_1,\ldots,d_k),num_{\mbox{eval}}]=1 \mbox{ if } k>nm \mbox{ and } d_k=d_{k-nm},$$ i.e. if we have evaluated all designs resulting from changing one coordinate and not seen an improvement.

 \begin{table}
 \small
\begin{tabular}{ p{3.9cm} l p{2.1cm}p{2.6cm}p{2.6cm}p{2.9cm}}
 Example &n& Number of automorphisms & Evaluations (CD) & Evaluations (ES)& Efficiency of design found with CD \\
 1. Small social network& 10 & 8 & 77 & 236 & 1 \\
2. Small social network &  10 & 1 & 145 & 511 & 0.944 \\
3. Larger social network & 20 &  8 & 127 & 221183 & 0.989 \\
4. Block design with neighbour effects & 12 & 384 &14 & 18766 & 0.873\\
5. Non-rectangular field trial& 15 & 2 & 82 & 1581572 & 0.931 \\
6. Crosover trial with dropouts& 15 & 6 & 93 & 90455& 1 \\ 
\end{tabular}\caption{\label{tab:CE}Comparison of coordinate descent (CD) and exhaustive search (ES) algorithms, both methods using automorphisms.}
\end{table}

We perform 100 random starts to mitigate the algorithm finding local maxima. To demonstrate how this algorithm can work with automorphisms, we apply this algorithm to the same 6 networks as in Table \ref{tab:JRSS2016} above. We present the results as Table \ref{tab:CE}. We can see that even this simple algorithm is compatible with the network framework of taking automophisms into account, and a very small number of function evaluations are made, with generally highly efficient designs being found.

 \section{Block designs as networks}
\label{sec:BlockDesigns}
We seek to generalise the framework developed to include a wider class of models. We start with a simple example.

Suppose we have four blocks, each of four experimental units. Let us suppose we wish to allocate treatments labelled $\{1,\ldots,m\}$ with corresponding unknown effects $\{\tau_1,\ldots,\tau_m\}$ to the experimental units to maximise some optimality criterion; for example we may wish to estimate 
$$\frac{2}{m(m-1)}\sum_{j=1}^{m-1}{\sum_{l=j+1
}^m{\var({\widehat{\tau_j-\tau_l})}}}.$$ This is defined to be $A_s$-optimality for
estimating the
differences in the treatment effects. 
We call this criterion function $\phi_1$.

A typical schematic for the structure of the experiment might be written down as in the left of Figure \ref{fig:BlockDesign}. It is clear that experimental units 1,2,3, and 4 are contained within block 1, and so on. We might choose to represent the experiment pictorially as in the right of Figure \ref{fig:BlockDesign}.

\begin{figure}
 \begin{minipage}[c]{0.4\linewidth}
\par\vspace{0pt}
\centering

\begin{tabular}{c | c c c c}

Block& \multicolumn{4}{c}{Experimental Units}\\
\hline
1&1 & 2 & 3 & 4\\
2&5 & 6 & 7 & 8\\
3&9 &10 &11 &12\\
4&13&14&15&16
\end{tabular}

\centering
\par\vspace{0pt}
  \end{minipage}%
  \begin{minipage}[c]{0.6\linewidth}
  \par\vspace{0pt}
  \centering
 \includegraphics[width=\linewidth]{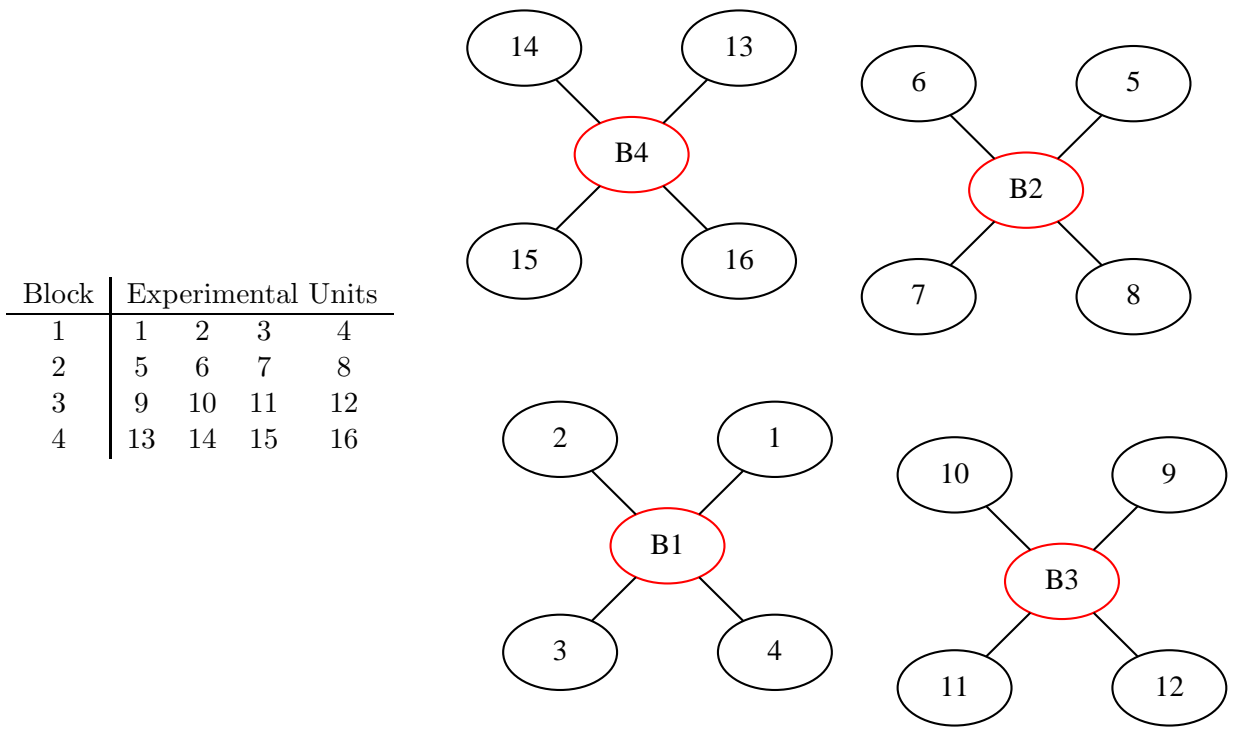}
\par\vspace{0pt}
\end{minipage}
\caption{Left: A traditional representation of a simple block design with 16 units in four blocks. Right: A representation of a blocked experiment as a network design. Experimental units numbered 1 to 16 are in black circles, and blocks are indicated by $B1, B2, B3, B4$ with red circles.}
\label{fig:BlockDesign}
\end{figure}

Why is this useful? Let us now imagine that the 20 nodes $\{1,\ldots,16,B1,B2,B3,B4\}$ are experimental units that are connected according to the relationship shown in Figure \ref{fig:BlockDesign}. We forget that the nodes have any special meaning now, save that we may assign treatments $1,\ldots,m$ only to the first 16 nodes, and that node B1 is assigned treatment $m+1$, node B2 assigned treatment $m+2$, B3 assigned $m+3$ and finally B4 assigned $m+4$.

In order to find the $A$-optimal design, we can write the model for our blocked experiment as  
\begin{equation}
Y_i= \mu+\tau_{t(i)} +\sum_{k=\{1,\ldots,16,B1,B2,B3,B4\}}{A_{ik}\gamma_{t(k)}}+\epsilon_i, \quad i=1,\ldots,16
\end{equation}
where A is the adjacency matrix where $A_{ij}=1$ whenever there is a link as shown in the Figure (e.g. $A_{14,B4}=1$, and $A_{ij}=0$ otherwise).

By writing $b_{j(i)}$ as $\sum_{k=\{1,\ldots,16,B1,B2,B3,B4\}}{A_{ik}\gamma_{t(k)}}$, we can see immediately that this is equivalent to
$$Y_i= \mu+\tau_{t(i)} +b_{j(i)}+\epsilon_i,$$ a more familiar representation of a blocked experiment, where $b_{j(i)}$ is block effect of experimental unit $i$ in block $j$. As $A_{ij}=1$ if and only if node $i$ is linked to node $j$, and this only happens when experimental unit $i$ is in block $j$ for $j=\{B1,B2,B3,B4\}$, we can  replace $\gamma_{t(B1)}=\gamma_{m+1}=b_{1}$, $\gamma_{t(B2)}=\gamma_{m+2}=b_{2}$, and so on. 

Perhaps more simply, the network effects of the special nodes representing blocks in the linear network effects model are replaced by the block effects.  We can think of the block effects as propagating like network effects from the special nodes B1, B2, B3, B4. We will normally wish to estimate some function of the treatment effects only (block effects are often not of interest but it is important to account for them in the design), and as the treatment effects for nodes B1, B2, B3, B4 are irrelevant (we cannot measure the experiment units B1, B2, B3, B4 directly as they are not real units!), we ignore them and use the same optimality criterion as before. As we are not estimating all effects, strictly this is now $A_S$ optimality rather than A optimality.

The properties of the two designs can be summarised in Table \ref{tab:augmented} .
\begin{table}
\begin{tabular}{c | p{6cm}  | p{6cm}}
&Original Problem & Network Problem\\
\hline 
No of Treatments & $m$ & $m+4$\\ \hline
Experimental Units &  $\{1,2,\ldots 16\}$& $\{1,2,\ldots,16, B1,B2,B3,B4\}$\\ \hline
  Wish to estimate &  Average pairwise variance of $\tau_i-\tau_j$ for all $1\le i<j\le m$&  Average pairwise variance of $\tau_i-\tau_j$ for all $1\le i<j\le m$\\ \hline
  Optimality criterion &  A &  $A_s$\\ \hline
  Restrictions &  Can apply any treatment to any unit. &  Can apply treatments 1,\ldots,m to units $\{1,2,\ldots,16\}$, and treatments 3,4,5,6 to units B1-B4.
\end{tabular}\caption{\label{tab:augmented}Comparison of original and augmented network design for $m=2$ treatments.}
\end{table}

We have thus shown (at least for this simple example) an equivalence between a traditional design (in this case a block design) and a network design as found in the paper \cite{Parker2016}. A major advantage of writing a block design in this way is that we can now find optimal designs for block designs in the same way as for network designs, and that we can use the automorphisms found from the network representation of the experimental framework in order to find optimal designs faster. The network setting allows us to develop one set of effective algorithms, rather than regarding blocked experiments as a special class  with its own method for finding optimal designs.

\subsection{Other block designs as networks}

 \begin{figure}
 \begin{minipage}[c]{0.4\linewidth}
\par\vspace{0pt}
\centering
\begin{tabular}{c|c|c|c}
&C1&C2&C3\\
\hline
R1&1&2 & 3 \\
R2&4&5&6\\
R3&7&8&9
\end{tabular}

\centering
\par\vspace{0pt}
  \end{minipage}%
  \begin{minipage}[c]{0.6\linewidth}
  \par\vspace{0pt}
  \centering
\includegraphics[width=\linewidth]{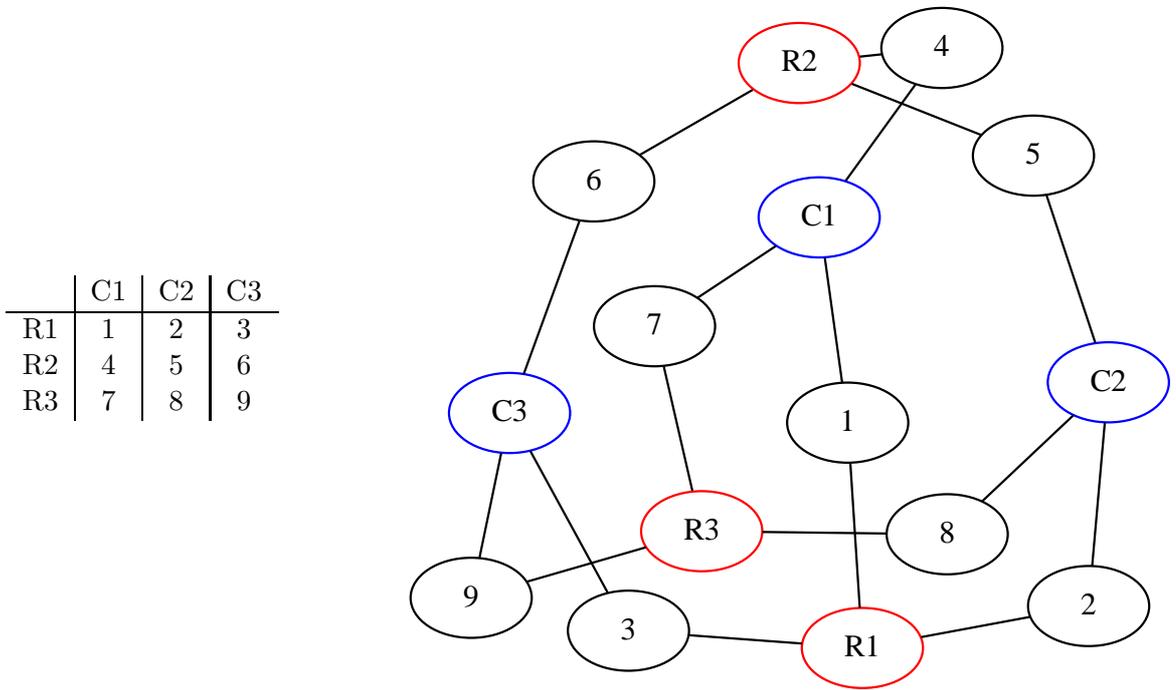}
\par\vspace{0pt}
\end{minipage}
\caption{A Row Column Design in (left) a traditional representation and (right) a network diagram.}
\label{fig:rowColumnDesign}
\end{figure}

In a similar method to the one-way block design, it is possible to represent a variety of block designs as networks. For all blocking factors each block can be written as a new network node, linked to all experimental units within that block. We present examples for a double-blocked experiment (a row-column design) as Figure \ref{fig:rowColumnDesign}. This is readily extendable to more than two blocking factors.

\begin{figure}
 \begin{minipage}[c]{0.4\linewidth}
\par\vspace{0pt}
\centering

\begin{tabular}{c c| c c c}
& &\multicolumn{3}{c}{Period} \\

\multirow{4}{*}{}& &P1 & P2& P3  \\  
\hline
\multirow{4}{*}{Subject}&a & 1& 2 & 3 \\
&b&  4 & 5 & 6\\\
&c&  7 & 8 &9\\
\end{tabular}

\centering
\par\vspace{0pt}
  \end{minipage}%
  \begin{minipage}[c]{0.6\linewidth}
  \par\vspace{0pt}
  \centering
 \includegraphics[width=\linewidth]{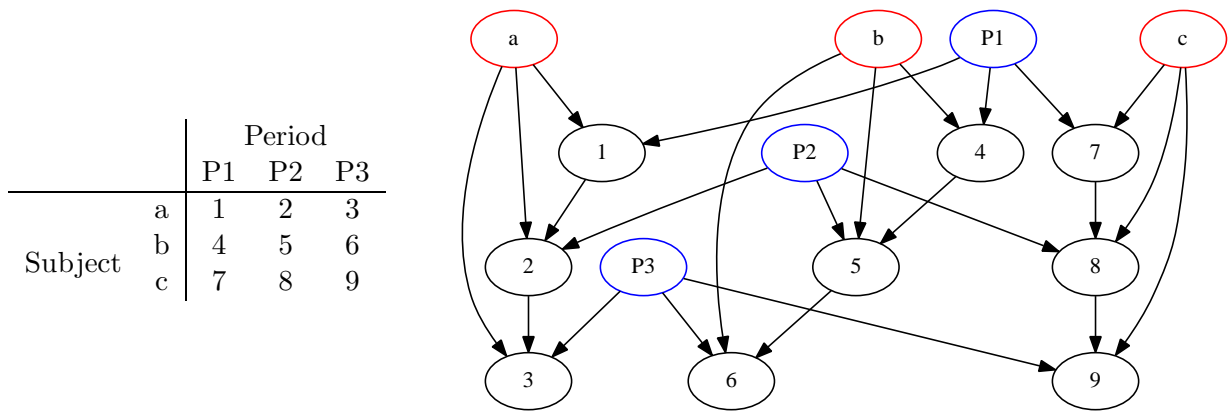}
\par\vspace{0pt}
\end{minipage}
\caption{A Crossover Design in (left) a traditional representation and (right) a network diagram.}
\label{fig:CrossoverDesign}
\end{figure}

In a crossover design, subjects receive treatments sequentially over several time periods, and an experimenter wishes to account for the treatment possibly still having some effect in a later time period than that for which it is applied. We can represent a crossover-design similarly to a row-column design as Figure \ref{fig:CrossoverDesign}. Experimental units are subject-period combinations, and we assume that subjects behave similarly, so subject is a blocking factor, and periods may be a blocking factor, so we add extra block nodes as we did for the row-column designs. In addition, we assume a treatment applied to a subject may affect the subject in the next time period, so there may be a carryover effect from experimental unit 1 to 2, 2 to 3, 4 to 5, etc. We represent this carryover effect by the network effect of the treatment given to the experimental unit preceding in time, but note this carryover effect only extends to the next time period, so links in the network are directed. This is similar to Example 6 in \cite{Parker2016}.

\subsection{Advantages of changing design space}
The argument explained in \ref{sec:rationale} applies: by evaluating only one design from each class of automorphic designs, we can make significant savings in the amount of time to evaluate candidate design. 

We evaluate several experimental structures to demonstrate this in practice, for which the experimental designs are well-known:
\begin{enumerate}
 \item 3 blocks of size 3 ($n=9$), with 3 treatments. (The optimal designs are randomised complete block designs.)
 \item 4 blocks of size 3 ($n=12$), with i) 3 and ii) 4 treatments. (The optimal designs are i) randomised complete block designs and ii) balanced incomplete block designs.)
 \item A row-column structure with 3 rows and 3 columns, each row-column intersection containing a single experimental unit, with 3 treatments. (The optimal designs are Latin Squares of size 3.)
 \item A row-column structure with 4 rows and 4 columns, each row-column intersection containing a single experimental unit, with i)3 and ii)4 treatments. (The optimal designs in ii) are Latin Squares of size 4.)
\end{enumerate}

\begin{table}
\small
\begin{tabular}{p{3.5cm} l l p{2cm}p{2cm}p{2cm}p{2cm}p{2cm}}
 Example &n&m& Number of automorphisms & Evaluations without automorphisms & Evaluations with automorphisms & Time without automorphisms & Time with automorphisms\\
 1. 3x3 Blocks& 9 & 3&1296  & 2925 & 94& 2.52 & 1.54\\
2i. 4x3 Blocks &  12 & 3&82944 & 86126& 379 & 55.44 & 310.02\\
2ii. 4x3 Blocks &  12 & 4&82944& 605960  & 1808& 378.82 & 1051.54\\
3. 3x3 Row Column & 9 & 3&  241 & 72 & 2807 & 1.9 & 0.48\\
4i. 4x4 Row Column & 16 & 3 &1152 &7123656 &34873& 6051.12 &493.32\\
4ii. 4x4 Row Column & 16 & 4&1152  & 170863644 &1610909& 141456.6 & 14123.94\\
\end{tabular}\caption{\label{tab:blockDesigns} Designs for various block designs. Number of evaluations of optimality criterion and computational time with and without automorphisms}
\end{table}

We do not claim that these designs would be sensibly found via this method, as the solutions are known analytically, but we seek to demonstrate the benefits of using automorphisms in reducing computational time.

The results are presented as Table \ref{tab:blockDesigns}. Clearly the number of optimality function evaluations is vastly reduced for networks a large number of automorphisms. However, for some networks, such as 2i and 2ii, the overhead in the implementation caused by searching for automorphisms and then checking the lexicographical order of the candidate designs might make the use of automorphisms inefficient. 

For larger designs, such as 4i, with a moderate number of automorphisms, we see a more than tenfold reduction in processing time and a hundred-fold reduction in number of evaluations, suggesting that this method works better for networks with a moderate number of automorphisms. 

In practice, by taking into account network structure, practitioners may be able to make a very large saving in time to find optimal designs with very little modification to existing algorithms, meaning designs found using stochastic algorithms in fixed computing time may be better.

\subsection{Computational Justification}
\label{sec:complexity}
Recall that $|\desX|=n^m$ for unstructured treatments. For our $A_s$ optimality criterion, and indeed most common optimality criteria, when calculating $f(x)$ we must calculate the Fisher information matrix and invert it to find a variance-covariance matrix, then make some calculation of this final matrix. We assess the typical computational complexity for each design we consider.
\begin{itemize}
 \item The complexity for calculation of the Fisher information matrix from $F^TF$ where $X$ is an $n\times(2m)$ matrix is typically $O(n^2m)$.
 \item The Fisher information matrix in our model (\ref{eq:linearNetworkEffects}) is of size $2m\times2m$. Inverting the matrix depends on the algorithm used, but is $O(m^k)$ where $2<k\le3$.
 \item Calculating the optimality criterion from the $2m\times2m$ variance-covariance matrix depends on what criterion is used; the trace (A-optimality) is $O(m)$, taking the determinant (D-optimality) will typically be $O(m^k)$ where $2<k\le3$.
 
\end{itemize}
  Thus for each design evaluated, and assuming a simple optimality criterion, as $n>>m$ (in general we have many more experimental units than treatments) the limiting step is the first step above and therefore calculating $f(x)$ has computational complexity of $O(n^2m)$.
  
  The complexity of the overhead in the new framework algorithm involves i) the initial time to calculate automorphisms originally, and ii) the computational cost of checking whether each design is lowest lexicographically amongst all possible automorphic designs. 
  
  Let us assume the size of $\texttt{isos}$ is $z$. i.e. there are $z$ automorphisms for our network. We map each design of length $n$ to an automorphism of itself which reorders the design, each time this operation is $O(n)$. We must do this $z$ times (once for each automorphism), and then sort the resulting list of automorphic designs to find the smallest, which can be done in $O(z\log z)$ for a good sorting algorithm. Thus the overall computational complexity of checking whether a design is first lexicographically is $O(nz + z\log z)$.

  Thus the computational complexity involved in the proposed algorithm where we first check that a design is valid, and if it is calculate the design, is $O(nz + z\log z+ n^2m/z)$, where the $z$ in the denominator in the second term arises because we must calculate $f(x)$ for one design in every $z$.
  
  Evaluating one design for the algorithm without automorphisms is $O(n^2m)$; for the new method is $O(nz+z\log z+ n^2m/z])$. The ratio of the new to the old is $$O\left(\frac{z}{nm}+\frac{z\log z}{nm}+\frac{1}{z}\right)\approx O\left(\frac{z\log z}{nm}+\frac{1}{z}\right)$$
  
  Thus the effectiveness of our new algorithm depends on the relative sizes of $n$, $z$, and $m$. We find $z$ must be small compared to $nm$ but not too small. This is supported by the results in Table \ref{tab:blockDesigns}, which show the computational time to be reduced significantly for moderate $z$, but actually increased for very large $z$, as the overhead in finding and ordering automorphisms is high here.

\section{Conclusions}
\label{sec:Conclusions}
We have shown that the use of automorphisms for reducing the number of evaluations required of an optimality criterion function is effective on designs where experimental units have a network structure. Moreover, we have shown that we can take block designs with no apparent network structure, such as one-way blocked experiments, row-column experiments, and crossover designs, and add block nodes to induce a network structure. Using automorphisms on these experiments with induced networks is also effective at reducing the complexity of experimental design algorithms.

From a practical point of view, many algorithms for design exist in isolation; we must program an algorithm for split-plot designs differently, perhaps, to how we program row-column designs. We argue that a framework such as we suggest with this paper may promote general purpose algorithms, neggating the need to maintain different algorithms for particular designs. Although algorithms for design can take into account automorphic designs to avoid recalculation of $f(x)$ for an automorphic design, in general they do not, and this network representation allows automorphisms to be found readily by algorithms that are quick and available in existing software. 
We believe that this work may lead to standardisation of algorithms across seemingly different classes of experiment.

\bibliographystyle{Harvard}
\bibliography{NetworkDesign.bib}

\begin{thebibliography}{}

\bibitem[Aral, 2016]{Aral2016}
Aral, S. (2016).
\newblock Networked experiments.
\newblock {\em The Oxford Handbook of the Economics of Networks}.

\bibitem[Babai, 2015]{Babai2015}
Babai, L. (2015).
\newblock Graph isomorphism in quasipolynomial time.
\newblock {\em CoRR}, abs/1512.03547.

\bibitem[Bailey, 2007]{Bailey2007}
Bailey, R. (2007).
\newblock Designs for two-colour microarray experiments.
\newblock {\em Journal of the Royal Statistical Society: Series C (Applied
  Statistics)}, 56(4):365--394.

\bibitem[Bailey and Cameron, 2011]{Bailey2011}
Bailey, R. and Cameron, P.~J. (2011).
\newblock Using graphs to find the best block designs.
\newblock {\em arXiv preprint arXiv:1111.3768}.

\bibitem[Bailey and Cameron, 2009]{Bailey2009}
Bailey, R.~A. and Cameron, P.~J. (2009).
\newblock Combinatorics of optimal designs.
\newblock {\em Surveys in Combinatorics}, 365(19-73):3.

\bibitem[{Basse} and {Airoldi}, 2017]{Basse2017}
{Basse}, G.~W. and {Airoldi}, E.~M. (2017).
\newblock {Preprint: Model-assisted design of experiments in the presence of
  network correlated outcomes}.
\newblock {\em ArXiv e-prints}.

\bibitem[Bulutoglu and Margot, 2008]{Bulutoglu2008}
Bulutoglu, D.~A. and Margot, F. (2008).
\newblock Classification of orthogonal arrays by integer programming.
\newblock {\em Journal of Statistical Planning and Inference}, 138(3):654--666.

\bibitem[Colbourn and Colbourn, 1981]{Colbourn1981}
Colbourn, M.~J. and Colbourn, C.~J. (1981).
\newblock Concerning the complexity of deciding isomorphism of block designs.
\newblock {\em Discrete Applied Mathematics}, 3(3):155--162.

\bibitem[Conte et~al., 2004]{Conte2004}
Conte, D., Foggia, P., Sansone, C., and Vento, M. (2004).
\newblock Thirty years of graph matching in pattern recognition.
\newblock {\em International journal of pattern recognition and artificial
  intelligence}, 18(03):265--298.

\bibitem[Cordella et~al., 2001]{Cordella2001}
Cordella, L.~P., Foggia, P., Sansone, C., and Vento, M. (2001).
\newblock An improved algorithm for matching large graphs.
\newblock In {\em 3rd IAPR-TC15 workshop on graph-based representations in
  pattern recognition}, pages 149--159.

\bibitem[Koutra, 2017]{Koutra2017}
Koutra, V. (2017).
\newblock {\em Designing Experiments on Networks}.
\newblock PhD thesis, University of Southampton.

\bibitem[Ma et~al., 2001]{Ma2001}
Ma, C.-X., Fang, K.-T., and Lin, D.~K. (2001).
\newblock On the isomorphism of fractional factorial designs.
\newblock {\em journal of complexity}, 17(1):86--97.

\bibitem[Meyer and Nachtsheim, 1995]{Meyer1995}
Meyer, R.~K. and Nachtsheim, C.~J. (1995).
\newblock The coordinate-exchange algorithm for constructing exact optimal
  experimental designs.
\newblock {\em Technometrics}, 37(1):60--69.

\bibitem[Parker et~al., 2016]{Parker2016}
Parker, B.~M., Gilmour, S.~G., and Schormans, J. (2016).
\newblock Optimal design of experiments on connected units with application to
  social networks.
\newblock {\em Journal of the Royal Statistical Society: Series C (Applied
  Statistics)}, 66(3):455--480.

\bibitem[Wit et~al., 2005]{Wit2005}
Wit, E., Nobile, A., and Khanin, R. (2005).
\newblock Near-optimal designs for dual channel microarray studies.
\newblock {\em Journal of the Royal Statistical Society: Series C (Applied
  Statistics)}, 54(5):817--830.

\end{thebibliography}

\section*{Appendix}

The following lists the edges for the six examples shown in this paper. These are also shown visually in examples in \cite{Parker2016}. ``\texttt{i-j}'' means an edge exists between $i$ and $j$, and the corresponding entry in the adjacency matrix $A(i,j)=A(j,i)=1$. If no edge it shown then $A(i,j)=A(j,i)=0$. Examples 1 to 5 are undirected networks. In Examples 6, the network is directed, such that ``\texttt{i->j}'' means that $A(i,j)=1$, but note that $A(j,i)\ne1$.

\begin{verbatim}
Example 1: 1-7, 2-7, 3-6, 4-5, 6-9, 9-10

Example 2: 1-5, 1-8, 1-10, 2-3, 2-4, 2-7, 2-8, 2-9, 2-10, 3-4, 3-7, 3-9, 
4-6, 4-7, 4-8, 4-9, 4-10, 5-9, 6-7, 7-10, 8-10, 9-10

Example 3: 1-2, 1-4, 2-3, 2-5, 2-9, 2-14, 2-17, 3-4, 3-8, 3-12, 3-13, 3-16,
4-6, 4-7, 4-9, 4-10, 4-11, 4-15, 4-20, 5-6, 5-7, 5-10, 5-14, 6-18, 7-19, 
9-11, 9-13, 9-16, 9-20, 10-15, 10-17, 10-18, 15-19

Example 4: 1-2, 2-3, 4-5, 5-6, 7-8, 8-9, 10-11, 11-12 

Example 5: 1-2, 1-4, 2-3, 2-4, 2-5, 3-5, 4-7, 4-8, 4-9, 5-9, 5-10,
6-7, 6-11, 6-12, 7-8, 7-11, 7-12, 7-13, 8-9, 8-12, 8-13, 8-14, 
9-10, 9-13, 9-14, 9-15, 10-14, 10-15, 11-12, 12-13, 13-14, 14-15, 

Example 6: 2->1, 3->2, 4->3, 6->5, 9->8, 10->9, 11->10, 13->12,
14->13, 15->14 
\end{verbatim}

\section*{Software}

A draft package to find designs for networks, including the augmented networks for block designs, is available at \url{https://www.dropbox.com/s/fgfn2f2my4yqlzr/networkDesign_0.0.0.9001.tar.gz?dl=0}. Scripts are provided to allow the reader to reproduce many of the results in this paper, as well as those in \cite{Parker2016}.

\end{document}